\documentclass[prl,aps,twocolumn,showpacs]{revtex4}
\usepackage{graphicx,color,psfrag,latexsym}
\usepackage{dcolumn}
\usepackage{amsmath,amssymb,epsf,bm}

\begin{document}

\title{Nonlinear Magneto-Optical and Magnetoelectric Phenomena in Topological Insulator Heterostructures}
\author{A.~G. Mal'shukov$^{1}$,
Hans Skarsv{\aa}g$^{2}$ and Arne Brataas$^{2}$}
\affiliation{$^1$Institute of Spectroscopy, Russian Academy of
Sciences, 142190, Troitsk, Moscow Oblast, Russia \\
$^2$Department of Physics, Norwegian University of Science and
Technology, NO-7491 Trondheim, Norway}
\begin{abstract}
Three-dimensional topological insulator films in contact with magnetic layers exhibit intriguing magneto-optical and magnetoelectric phenomena, but little is known beyond the linear response regime. We demonstrate that the presence of two electric fields parallel to the surface of the film induces a perpendicular AC polarization, whereas a static polarization can be induced by applying a single circularly polarized electromagnetic wave. These phenomena are determined by the second-order susceptibility, which
also controls the manner in which the anomalous Hall current is modulated by a perpendicular AC electric field. If the modulation frequency is low, the Hall effect is topologically protected and the nonlinear response vanishes.
\end{abstract}

\maketitle

Three-dimensional (3D) topological insulators (TIs) have unique physical properties and have attracted considerable attention because of their potential applications in many fields, including spintronics and quantum computations \cite{Qi RMP, Hasan}. These band insulators have gapless electronic surface states, which can be described by the massless two-dimensional (2D) Dirac-Weyl Hamiltonian at low energies \cite{Fu}. The surface states are robust with respect to perturbations that obey time-reversal symmetry; the energies of the surface states are linear in momentum, and the strong spin-orbit coupling rigidly locks the directions of the spin and momentum. If the surface of a TI is brought into contact with a magnetic layer, the time-reversal symmetry is broken and the metallic state of the 2D gas of surface electrons is transformed into a TI state. This state is characterized by a half-quantized anomalous Hall effect \cite{AHE}, which is closely related to the topological  magnetoelectric effect (TME) in the host TI \cite{Qi Basics TI,Essin}. The TME originates from the so-called axion term \cite{Qi RMP} in the Lagrangian of the electromagnetic field. This term is represented by the scalar product of the electric and magnetic fields multiplied by a topologically invariant constant. In addition to facilitating the quantized anomalous Hall effect, the TME appears in various transport, magnetostatic, and electrostatic phenomena on the surfaces of TIs \cite{Qi RMP,Qi Basics TI,TME}. Optical measurements are also useful in studying the TME. For example, the axion term leads to a rotation of the polarization of an electromagnetic wave incident on the TI surface, and theory \cite{Qi Basics TI, Faraday} predicts that the Faraday rotation angle is a universal number determined by the fine structure constant at frequencies below the terahertz range.

To date, most of the explored magnetoelectric effects have been limited to the linear response regime with respect to electric charges and currents induced by an electromagnetic field. Nonlinear magnetoelectric effects in TIs have not yet been addressed. To bridge this gap, we consider a family of nonlinear magneto-optical and magnetoelectric AC phenomena in a heterostructure composed of TI and magnetic layers. We calculate the corresponding second-order response, which originates from the axion \cite{Qi RMP} term in the electromagnetic action of the TI. Based on these calculations, we predict that a single circularly polarized or two linearly polarized electromagnetic waves incident on the surface of a TI will induce an electric polarization that is perpendicular to the interfaces. Surprisingly, the nonlinear susceptibility controlling this optoelectronic effect is closely related to the susceptibility determining variations in the Hall conductivity due to the presence of electric fields perpendicular to the interfaces. We demonstrate that the latter susceptibility has a specific frequency dependence associated with the topological nature of the TME. Therefore, the second-order electromagnetic response is expected to provide new, valuable and rich information about TME-related properties of TIs. Furthermore, because the electromagnetic pump, probe, and combined frequencies can be in resonance with the quantum transitions between the Dirac bands of the film, the nonlinear magnetoelectric and magneto-optic effects considered in this article are powerful tools for studying TIs. The predictions made in this work are important for the development of optoelectronic devices that incorporate TIs or topological superconductors.

Intuitively, a nonlinear variation of the TME could be achieved from a nonadiabatic modulation of the topological number in the Lagrangian of the TME caused by a time-dependent electric field because this number cannot be varied adiabatically. Indeed, we will demonstrate that such a modulation can be provided by an electric field $E_z$ that is perpendicular to the surface of the film. This field can be created by gates or electromagnetic irradiation. In turn, induced by $E_z$, the temporal variations of the axion field lead to time-dependent nonlinear corrections to the magnetoelectric and magneto-optical phenomena in TIs, such as the Hall effect, Faraday/Kerr rotation, and the magnetic monopole, which appears as an image of the charge near the surface of the TI \cite{TME}. However, not all of the nonlinear effects discussed here can be reduced to this simple picture. In general, the nonlinear term in the Lagrangian depends on three fields that are coupled to each other through a temporally nonlocal, nonlinear susceptibility. The nonlinear coupling gives rise to very different physical phenomena depending on whether these fields are static or AC driving fields. Nevertheless, in some cases, the seemingly different phenomena can be represented by the same second-order nonlinear susceptibility. Therefore, we will consider a subset of these phenomena, which are depicted in Fig. 1 and categorized into two proposed scenarios:

{\it Scenario 1}. As shown in Fig. 1a, two electromagnetic waves with frequencies $\Omega_1$ and $\Omega_2$ and with noncollinear electric fields directed parallel to the surface of the film induce oscillations of the electric polarization in the $z$-direction at the frequencies $\Omega_1\pm \Omega_2$. If $\Omega_1 = \Omega_2$, the induced polarization contains a static component. Such an optical rectification effect is also the result of the second-order response to a single circularly polarized wave. It will be shown that this rectification effect can be expressed through the Faraday rotation angle. Therefore, these effects are related.

{\it Scenario 2}. Consider the anomalous Hall current $j_y$ driven by an electric field $E_x(\Omega)$. There is also a perpendicular electric field $E_z(\Omega^{\prime})$, as shown in Fig. 1b. The anomalous Hall current driven by the electric field $E_x$ is also modulated in time by the $E_z$-field. If $E_x$ is static, the modulated current $j_y$ only varies with the frequency of the perpendicular electric field 
$\Omega^{\prime}$. If the system is in the TI state with a topological (quantized) Hall effect, the Hall conductance is robust and is unaffected by an adiabatic change in $E_z$. Therefore, the nonlinear current vanishes in the adiabatic limit $\Omega^{\prime} \rightarrow 0$. However, with increasing frequency $\Omega^{\prime}$, resonant transitions between gapped surface states lead to a strong enhancement in the nonlinear magnetoelectric susceptibility. In ultrathin TI films \cite{exp thin layers}, this enhancement allows one to detect the topological phase transition between the Hall insulator and topologically trivial phases, in which the gap and thus the resonance frequency vanish.

\begin{figure}[tp]
\includegraphics[width=8cm]{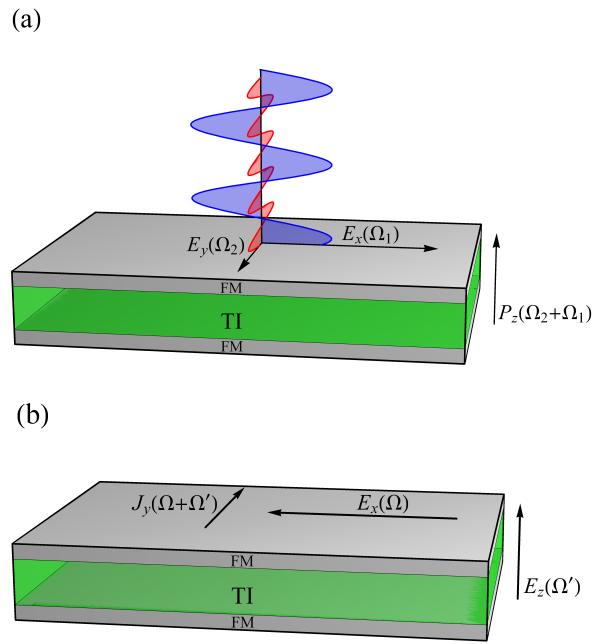}
\caption{(Color online) A topological insulator (TI) sandwiched between magnetic insulators (FM). a) Two linearly polarized electromagnetic waves $E_x(\Omega_1)$ and $E_y(\Omega_2)$ induce the AC transverse polarization $P_z(\Omega_1 +  \Omega_2)$. If
$\Omega_1=-\Omega_2$, the transverse polarization induced by two waves or by one circularly polarized wave is static. b) The AC in-plane $E_x(\Omega)$ and transverse $E_z(\Omega^{\prime})$ electric fields 
give rise to the nonlinear anomalous Hall current $J_y(\Omega + \Omega^{\prime})$. $E_z(\Omega^{\prime})$ can be induced by gates or by an electromagnetic wave.  }
\label{fig1}
\end{figure}

These phenomena and other nonlinear optical effects \cite{twophoton} can be studied in epitaxially grown ultrathin TI films and superlattices \cite{exp thin layers}. Ab initio calculations \cite{calculations for superlattices} for superlattices composed of Bi$_2$Se$_3$ and magnetic insulator layers have demonstrated the high tunability of system parameters, which is important for potential applications of these materials and for 
observing nonlinear magnetoelectric effects. The Fermi level position and compensation of bulk carriers in TI films can be adjusted through gate manipulation and through appropriate doping \cite{gate control}. As a model system, we consider an ultrathin TI film with magnetic layers on both the top and bottom surfaces. The film is assumed to be sufficiently thin that tunneling can occur between the Dirac states of the top and bottom surfaces. An alternative setup includes a thin magnetic layer between two TI films. In this case, the Dirac states on the TI surfaces in contact with the magnetic layer are tunnel-coupled to each other through this layer. Such a coupling can be stronger than that in the former case because the quintuple layers in Bi$_2$Se$_3$ are weakly coupled to each other through Van der Waals forces; this weak coupling follows from calculations of Bi$_2$Se$_3$/MnSe superlattices \cite{calculations for superlattices}. In this case, the two TI films adjacent to the magnetic layer are not necessarily very thin, and the couplings to the other surface states of the films can be neglected. Both setups are described by the same low-energy effective two-layer Hamiltonian \cite{calculations for superlattices,film Hamiltonian,two minima}:
\begin{equation}\label{H}
H=v(\mathbf{k}\times\bm{\sigma})_z\tau_3 + \Delta\tau_1 +m\sigma_z -e\phi(t)\tau_3\,,
\end{equation}
where the Pauli matrices $\tau_1,\tau_2,\tau_3$ and the vector $\bm{\sigma}\equiv[\sigma_x,\sigma_y,\sigma_z]$ correspond to the pseudospin and electron spin, respectively; $\Delta$ is the tunneling parameter; $m$ is the exchange energy; and $\phi(t)\equiv\phi+\delta\phi(t)$ denotes the electric potential difference between the TI surfaces. The electric potential difference can be induced by external gates and by structure inversion asymmetry and consists of time-independent and time-dependent parts. The latter is a small perturbation. The electric field parallel to the surface is represented by the vector potential $\mathbf{A}(t)$ that contributes to the interaction Hamiltonian $-(ev/c)(\mathbf{A}(t)\times\bm{\sigma})_z\tau_3$. Below, we set $e=c=\hbar=1$.

The eigenenergies of the Hamiltonian (\ref{H}) for a static perpendicular electric field $\delta\phi(t)=0$ and for $\mathbf{A}=0$ are
\begin{equation}\label{E}
E_{\mathbf{k}}=\pm\sqrt{v^2k^2+m^2+R^2 \pm 2\sqrt{m^2R^2+v^2k^2\phi^2}}\,,
\end{equation}
where $R^2=\Delta^2+\phi^2$. We consider the case in which $\phi^2<|m|R$, so the maxima (minima) of all four energy bands are centered at $k=0$. If $R=|m|$, the gap between the lowest-energy electron and hole bands closes, and the bands transform into the Dirac cone with a particle velocity $\tilde{v}=v\Delta/|m|$. In contrast, the system is a TI characterized by anomalous Hall conductance if $R<|m|$, whereas the system is an insulator with zero Hall conductance if $R>m$ \cite{Qi Basics TI}.

Let us consider scenario 1 with AC electric fields $\mathbf{E}_1(\Omega_1)$ and $\mathbf{E}_2(\Omega_2)$ parallel to the surface of the film. We will demonstrate that these fields induce the transverse polarization $P_z(\Omega_1+\Omega_2)$, which arises from the surface charge $\rho_s$ that accumulates on the top and bottom film surfaces.
In the geometry of the film, this polarization is $P_z=(\rho_{s~\text{top}}-\rho_{s~\text{bottom}})/2$, so the one-particle operator of the transverse polarization is $\tau_3$. Second-order perturbation theory, in response to $\mathbf{E}_1$ and $\mathbf{E}_2$, yields
\begin{equation}\label{Pz}
P_z(\Omega_1+\Omega_2)=
-\frac{i}{\Omega_2}\tilde{\chi}_{yx}(\Omega_1,-\Omega_1-\Omega_2)\left(\mathbf{E}_1\times \mathbf{E}_2 \right)_z \,,
\end{equation}
where the nonlinear susceptibility $\tilde{\chi}_{yx}(\Omega,\Omega^{\prime})$ can be expressed in terms of Keldysh Green’s functions as
\begin{eqnarray}\label{jy}
&&\tilde{\chi}_{yx}(\Omega,\Omega^{\prime})=-\frac{1}{2\Omega}\sum_{\mathbf{k}}\int \frac{d\omega}{2\pi}\times \nonumber   \\ &&\mathrm{Tr}[G_{\mathbf{k}}(\omega-\Omega-\Omega^{\prime})\Gamma_y G_{\mathbf{k}}(\omega)\Gamma_x G_{\mathbf{k}}(\omega-\Omega)\Gamma_z +\nonumber   \\
&&\Gamma_z
G_{\mathbf{k}}(\omega+\Omega)\Gamma_x G_{\mathbf{k}}(\omega)\Gamma_y  G_{\mathbf{k}}(\omega+\Omega+\Omega^{\prime})]^K \,,
\end{eqnarray}
where $\Gamma_x=-v\sigma_y\tau_3$, $\Gamma_y=v\sigma_x\tau_3$, and $\Gamma_z=\tau_3$. The superscript "K" denotes the Keldysh component of the products of the Green functions, and the trace is the sum over the spin and pseudospin variables. The retarded ($G^r$) and advanced ($G^a$) functions are $G^{r/a}=(\omega-H\pm i\delta)^{-1}$. At equilibrium, the Keldysh function is $G^K=(G^{r}-G^{a})\tanh(\omega/2k_B T)$. These functions are combined into a 2$\times$2 matrix $G_{11}=G^r$, $G_{22}=G^a$, $G_{12}=G^K$, and $G_{21}=0$ \cite{Keldysh}. The frequency 
integrals from the two expressions in the integrand are identical, which 
can be illustrated by transforming one expression into the other through the unitary transformation $\sigma_x\tau_2 H\sigma_x\tau_2=-H^{\mathrm{T}}$, $\tau_2 \sigma_x\Gamma_i\sigma_x\tau_2=-\Gamma^{\mathrm{T}}_i$, $i=x,y,z$, and $\sigma_x\tau_2 G^{r/a}_{\mathbf{k}}(\omega)\sigma_x\tau_2=-G^{a/r\mathrm{T}}_{\mathbf{k}}(-\omega)$, where $\mathrm{T}$ denotes a transposed matrix. Considering that  $\Gamma_{x/y}=-\partial H/\partial k_{x/y}$, we simplify Eq. (\ref{jy}) to
\begin{eqnarray}\label{jy2}
&&\tilde{\chi}_{yx}(\Omega,\Omega^{\prime})=-(\Omega+\Omega^{\prime})\sum_{\mathbf{k}}\int \frac{d\omega}{2\pi}\times \nonumber   \\ &&\mathrm{Tr}\left[ \frac{\partial G_{\mathbf{k}}(\omega-\Omega-\Omega^{\prime})}{\partial k_y} G_{\mathbf{k}}(\omega)\frac{\partial G_{\mathbf{k}}(\omega-\Omega)}{\partial k_x} \Gamma_z \right]^K\,,
\end{eqnarray}
\begin{figure}[tp]
\includegraphics[width=9cm]{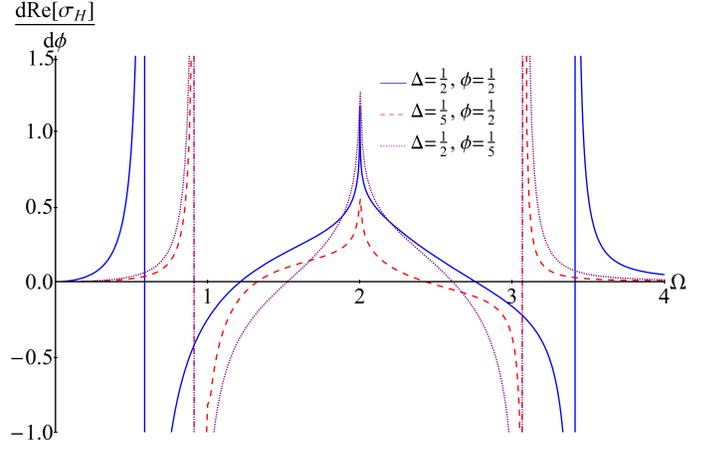}
\caption{(Color online) Derivative with respect to the potential of the real part of the Hall conductivity. If $\Omega$ is less than the lowest frequency peak, the derivative coincides with $\tilde{\chi}_{yx}(\Omega,0)$ in Eq. (\ref{Pz}). $\Omega$ and the other parameters are given in units of $m$.  }
\label{fig2}
\end{figure}

Based on Eq.~(\ref{Pz}), AC polarizations with frequencies $\Omega_1\pm\Omega_2$ are produced by two fields $\mathbf{E}_{1(2)} \sim (\mathbf{E}_{1(2)}(\Omega_{1(2)})\exp(i\Omega_{1(2)} t)+c.c.) $ when their frequencies differ. The polarization 
contains a static component if $\Omega_1=-\Omega_2$. The static polarization can also be induced by single circularly polarized electromagnetic wave. In this case, $\mathbf{E}_1=\mathbf{E}_2^*$ and 
$\left(\mathbf{E}_1\times \mathbf{E}^*_1 \right)_z\neq 0$. The static $P_z(0)$ is determined by $\tilde{\chi}_{yx}(\Omega_1,0)$. The nonlinear susceptibility $\tilde{\chi}_{yx}(\Omega_1,0)$ is closely related to the frequency-dependent Hall conductivity $\sigma_H(\Omega_1)$. This relation follows from the known correspondence between the optical rectification effect and linear electro-optical effect \cite{Landau}. In our case, this correspondence implies that if the frequency is smaller than the gap, $\Omega_1 < E_g$, $\tilde{\chi}_{yx}(\Omega_1,0)=d\sigma_{H}(\Omega_1)/d\phi$. Fig. 2 displays the frequency dependence of $d\text{Re}[\sigma_{H}(\Omega)]/d\phi$. As expected,
$\tilde{\chi}_{yx}(\Omega,0)$ vanishes when $\Omega \rightarrow 0$ due to the topological protection of the static Hall effect with respect to adiabatic variations in $\phi$. The peaks at $2|m+R|$, $2|m-R|$, and $(|m+R|+ |m-R|)$ correspond to interband transitions.

In general, the nonlinear susceptibility $\tilde{\chi}_{yx}(\Omega,\Omega^{\prime})$ has several resonances, which are determined by transitions between the surface bands (\ref{E}). As an example, let us consider oscillations of $P_z$ for frequencies below the terahertz range; these oscillations are induced by two infrared fields. We assume that one of these fields, for example, $\mathbf{E}_{1}$, is close to the resonance frequency for the quantum transition from the top valence band, with energy $-E^{(2)}_{k}$, to the upper conduction band $E^{(1)}_{k}$. Simultaneously, the other field, $\mathbf{E}_{2}$, is assumed to be in resonance with a transition between the upper and lower conduction bands. In summary, we assume that $\Omega_1 = (E^{(2)}_{0}+E^{(1)}_{0}) - \delta$ and $\Omega_1- \Omega_2= 2E^{(2)}_{0} - \delta^{\prime}$, where $\delta$ and $\delta^{\prime}$ are small frequency detunings from the resonance conditions. In this case, the induced polarization oscillates at a frequency close to $2E^{(2)}_{0}$. This frequency can vary with $\phi$ and vanishes at the topological phase transition in which $E^{(2)}_{0}=0$. Assuming that the incident electromagnetic waves $\mathbf{E}_{1}$ and $\mathbf{E}_{2}$ are linearly polarized in the $x$ and $y$ directions, at small detunings, one obtains from Eq. (\ref{Pz})

\begin{equation}\label{Pz2}
P_z=\frac{ie^3}{2\pi}\frac{\phi^3}{R^3}\frac{m(2\beta-\alpha)}{16|m|^3}\frac{\partial}{\partial \beta}\left(\frac{\ln|\beta\delta^{\prime}/\alpha\delta|}{\alpha\delta-\beta\delta^{\prime}}\right)
E_{1x}E^*_{2y}\,,
\end{equation}
where $\alpha=2[\partial E^{(2)}/ \partial X]_{k=0}$, and $\beta=[\partial( E^{(2)}+E^{(1)})/ \partial X]_{k=0}$ with $X=v^2 k^2$. At $\delta\sim\delta^{\prime}$, the resonance enhancement is $\sim\delta^{-1}$. If $\Delta \sim m \sim \phi$, we obtain $|P_z|\sim e^3|E_{1x}E^*_{2y}| /32\pi\delta\Delta$.

General symmetry relations for the second-order susceptibilities \cite{Landau} allow one to establish a correspondence between various nonlinear phenomena, similar to the manner in which we established a correspondence between the static polarization induced by a circularly polarized wave and $d\sigma_{H}(\Omega)/d\phi$. In turn, the latter derivative determines the effect of a static electric field on the Faraday rotation 
of a linearly polarized electromagnetic wave represented by the electric field $E_x(\Omega)$. In this manner, $\tilde{\chi}_{yx}(\Omega,\Omega^{\prime})$ from Eq. (\ref{Pz}) can be associated with a nonlinear effect that changes the anomalous Hall conductance by a time- dependent perpendicular electric field. The nonlinear Hall current density $j_y$ can be expressed in the form
\begin{equation}\label{chi}
j_y(\Omega+\Omega^{\prime})=
\chi_{yx}(\Omega,\Omega^{\prime})E_x(\Omega)\delta\phi(\Omega^{\prime}),
\end{equation}
where $\chi_{yx}=-\chi_{xy}$ is similar to Eq. (\ref{jy2}); however, there is the important difference that the operators under the trace sign are cyclically permuted, so the first operator in the product is $G_{\mathbf{k}}(\omega)$. The order of operators in the product is important in determining the proper Keldysh component associated with a calculated observable. In "transparent" systems, where $\Omega, \Omega^{\prime}$, and $|\Omega \pm \Omega^{\prime}|$ are below the energy gap, $\tilde{\chi}_{xy}(\Omega,\Omega^{\prime})=\chi_{xy}(\Omega,\Omega^{\prime})$, as follows from the symmetry relations for two-photon susceptibilities \cite{Landau}.
Let us now consider $\chi_{xy}(\Omega,\Omega^{\prime})$ for $\Omega \rightarrow 0$, whereas $\Omega^\prime$ remains finite. If $\Omega=0$, $\chi_{yx}$ vanishes as $\Omega^{\prime} \rightarrow 0$ because one cannot change the Hall conductivity, which is expressed by the first Chern number through an adiabatic variation of $\phi$. To calculate nonadiabatic corrections, we focus on the properties close to the topological phase transition at $R=|m|$. In this regime, the corresponding energy gap is small, $E_g=2||m|-R| \ll |m|$. If $\Omega^{\prime}< E_g$, the dominant contribution from the integration over $k$ in Eq. (\ref{jy2}) comes from the lowest-energy conduction bands and the highest-energy valence bands. At $k_BT\ll E_g$, in the first -order with respect to $E_g/m \ll 1$, we obtain for $\chi_{xy}(\Omega^{\prime})\equiv \chi_{xy}(0,\Omega^{\prime})$
\begin{equation}\label{sigmaH1}
\chi_{xy}(\Omega^{\prime})=\frac{e^2}{h}\frac{4\phi E_g}{m \Omega^{\prime 2}}\left(\frac{\Omega^{\prime 2}}{E_g^2-\Omega^{\prime 2}}-\frac{3E_g}{2\Omega^{\prime}}\ln\left|\frac{E_g+\Omega^{\prime}}{E_g-\Omega^{\prime}}\right|
+3\right)\,.
\end{equation}

Let us consider the low-frequency regime. By expanding Eq.~(\ref{sigmaH1}) in terms of small $\Omega^{\prime}/E_g$, we obtain
\begin{equation}\label{sigmaH2}
\chi_{yx}(\Omega^{\prime})=\Omega^{\prime 2}\frac{e^2}{h}\frac{8\phi}{5mE_g^3 }\,.
\end{equation}

As expected, the nonlinear Hall conductivity vanishes as $\Omega^{\prime} \rightarrow 0$ due to the topological constraint on adiabatic variations of $\sigma_H$. Let us illustrate this fundamental property of the quantum Hall conductivity by considering a case in which the electric field $E_x$ acts only on electrons on the top surface. The tunnel coupling between the top and bottom surfaces produces Hall currents on both surfaces. The sum of these currents can be expressed by the first Chern number. However, the currents are not independently invariant and can be varied adiabatically by changing $\phi$, as shown in Fig. 3.
A considerable enhancement in the nonlinear effect can be observed near the resonance $\Omega^{\prime}=E_g$. From Eq. (\ref{jy2}), we obtain the following in such a resonance regime:
\begin{equation}\label{sigmaH3}
 \chi_{yx}(\Omega)=\frac{2e^2}{h}\frac{m \Delta^2}{(|m|R-\phi^2)R^2} \frac{\phi}{E_g-\Omega^{\prime}}\,.
\end{equation}
When the gap is small, this expression coincides with Eq. (\ref{sigmaH1}) near the resonance. This result and Eq. (\ref{sigmaH1}) are valid only if $\Omega^{\prime} < E_g$. If $\Omega^{\prime} > E_g$, one must introduce a final dephasing time to obtain a convergent integral over $k$.

\begin{figure}[tp]
\includegraphics[width=9cm]{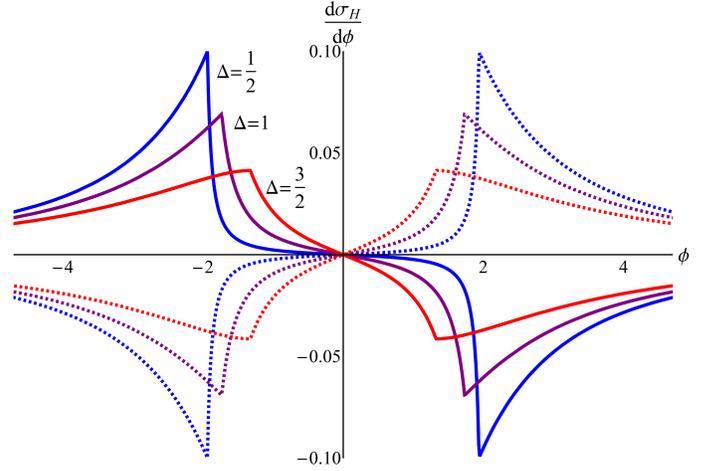}
\caption{(Color online) Variations in the anomalous Hall conductivity (in units of $e^2/h$) of the top (solid) and bottom (dashed) surfaces of the film as a function of the transverse potential for various tunneling parameters. $\phi$ and $\Delta$ are given in units of $m$.  }
\label{fig3}
\end{figure}

In conclusion, we considered a 3D TI thin film in contact with magnetic layers. We demonstrated that strong spin-orbit coupling combined with the Zeeman field, which is induced by the magnetic layers and which breaks the time-reversal symmetry, leads to a family of non-trivial magneto-optical and magnetoelectric phenomena. These phenomena arise from nonlinear coupling of the three 
electric fields $E_x$, $E_y$, and $E_z$. A time-dependent field $E_z$ leads to nonlinear oscillations of the anomalous Hall current. Due to the topological nature of this current, the amplitude of the nonlinear oscillations vanishes when the frequency $\Omega$ of $E_z$ decreases. Furthermore, the amplitude strongly increases when $\Omega$ is close to the gap between the valence and conduction surface bands. The three-field nonlinear coupling also leads to an unusual optical phenomenon: two linearly polarized electromagnetic waves, incident parallel to the film normal, induce an electric polarization directed along the normal. The polarization oscillates at a frequency that is the sum or difference of the frequencies of the incident waves. Additionally, static polarization can be created by a single circularly polarized wave. The nonlinear susceptibility that determines this polarization can be expressed through variations in the Faraday rotation angle caused by a static $E_z$.


\end{document}